\documentclass[twocolumn,showpacs,floatfix,preprintnumbers,amsmath,amssymb,prb]{revtex4}

\usepackage{graphicx}    
\usepackage{epsfig}      
\usepackage{dcolumn}     
\usepackage{bm}          

\def\sb#1{$_{#1}$}
\def\sp#1{$^{#1}$}
\def\etal{{\it et al.}}
\def\im3{{Im\overline 3}}

\def\lsmo{La$_{2-2x}$Sr$_{1+2x}$Mn$_2$O$_7$}
\def\gr{$G(r) = {2\over\pi}\int_0^{\infty}Q[S(Q)-1]\sin Qr\>dQ$}

\def\mn#1{Mn$^{#1+}$}

\def\nb#1{{\bf Xiangyun[#1]}}

\def\bibpathsjb#1{C:/billinge/research/papers/bib/#1}

\begin{document}

\title{Evidence for nano-scale inhomogeneities in bilayer manganites in the Mn\sp{4+} rich region: $\bm{0.54 \leq x \leq 0.80}$}

\author{Xiangyun Qiu and Simon J. L. Billinge}
\affiliation{Department of Physics and Astronomy and Center for
Fundamental Materials Research, Michigan State University, Biomedical Physical
Sciences, East Lansing, MI 48824-2320.}
\author{Carmen R. Kmety and John F. Mitchell}
\affiliation{Material Science Division, Argonne National Laboratory, \\ 9700 S Cass Avenue, Argonne, Illinois, 60439}

\date{\today}

\begin{abstract}
The atomic pair distribution function (PDF) technique is employed to probe the atomic local structural responses in naturally double layered manganites \lsmo\ in the
doping range $0.54 \leq x \leq 0.80$.
Our low temperature neutron powder diffraction measurements suggest the coexistence of two different Jahn-Teller (JT) distorted MnO$_6$ octahedra when its ordered magnetic structure crosses over from type A ($0.54 \leq x \leq 0.66$) to type C/C$^*$ ($0.74 \leq x \leq 0.90$) ordering.
At all doping levels at low temperature the doped holes reside predominantly in the plane of the
bilayer.
In the type A magnetic ordering regime, the e$_g$ electrons appear to be significantly delocalized
in the plane resulting in undistorted octahedra, while in type C/C$^*$ regime, elongated JT distorted
octahedra are apparent.  This is consistent with  the presence of inhomogeneous
coexisting delocalized and localized electronic states.  No evidence of macroscopic phase separation
has been observed.  Such nanoscale inhomogeneities may explain the magnetically frustrated
behavior observed in the spin disordered ``gap'' region ($0.66 \leq x \leq 0.74$).
\end{abstract}

\maketitle
\section{INTRODUCTION}
The double layered manganite series La$_{2-2x}$Sr$_{1+2x}$Mn$_2$O$_7$ shows a rich and interesting, but somewhat poorly understood, phase diagram.\cite{kubot;jpcs99,ling;prb00} It also appears to be a prospective candidate for future application by showing a large colossal magnetoresistance (CMR) effect.\cite{morit;nl96}
The interplay between spin, charge, and lattice degrees of freedom is of critical importance  to many transitional-metal oxides with perovskite-related structures, and their delicate interactions have been intensely studied both theoretically and experimentally.
The reduced dimensionality obtained by constraining the lattice degree of freedom in this bi-layered system facilitates the investigation of strong correlation between electron-lattice coupling and magnetic properties, proving it to be one of the most interesting  CMR manganites.

\begin{figure}[tb]
\includegraphics[width=0.26\textwidth, keepaspectratio=1, angle=-90 ]{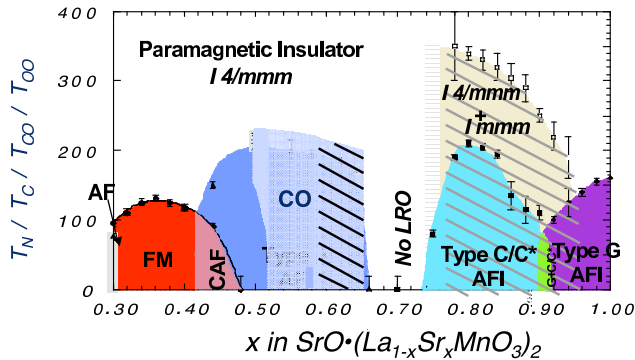}
\caption{Structural and magnetic phase diagram of the bilayer manganite \lsmo\ in the range 0.3 $\leq  x \leq 1.0$ determined by neutron
powder diffraction. Solid markers represent the magnetic transition temperature (T$_C$ or T$_N$); open squares delineate the tetragonal to orthorhombic
transition. Several magnetic phases are identified: ferromagnetic metal (FM), canted antiferromagnetic (CAF), and A-, C-, and G-type antiferromagnetic
insulators (AFI). The region marked "No LRO"  has no magnetic diffraction peaks at T $\ge 5$~K.
Samples in the region marked "CO" exhibit long-range charge ordering reflections in X-ray and/or electron diffraction. A temperature range
schematically indicated by the yellow square shows how this long-range charge-ordered state grows then disappears at low temperature. 
 \label{fig;phasedi}}
\end{figure}

Most studies focus on the Mn$^{3+}$ rich portion of the La$_{2-2x}$Sr$_{1+2x}$Mn$_2$O$_7$ phase diagram, where the important CMR phenomenon and the temperature driven Insulator-Metal (IM) transition are observed.\cite{morit;nl96} However, the phase diagram of the less studied Mn$^{4+}$ rich region shows many novel and intriguing properties.\cite{ling;prb00}
For convenience, this is reproduced in Fig.~\ref{fig;phasedi}.
For example, the type-A anti-ferromagnetic insulator (AFI) phase is the ground state in a wide range $0.46 \leq x \leq 0.66$ while it is more commonly found in cubic manganites at the lowest Sr$^{2+}$ dopings, though with some exceptions.\cite{morit;prb98,kuwah;prl99}  Also, a spin disordered ``gap'' region takes over for doping $x$ from 0.66 to 0.74; then a tetragonal to orthorhombic crystallographic  phase transition occurs sharply at $x$=0.74, followed by the type C/C$^*$ AFI phase up to 0.90 doping.  There is also  a wide charge ordered (CO) region ($0.48 \leq x \leq 0.66$).
These observations challenge the simple Goodenough-Kanemori (GK) rules\cite{goode;pr55} that successfully correlate the magnetic and structural properties of cubic perovskite manganites and are not well understood.

An A-type AFI phase, of ferromagnetic sheets antiferromagnetically coupled, at 50\% doping can be explained within the framework of the GK rules if the Mn 3$d_{x^2-y^2}$ orbitals are occupied by the $e_g$ electrons rather than the more commonly observed 3$d_{3y^2-r^2}$ occupancy. However, to explain the C/C$^*$ AFI phase (linear FM coupled chains that are antiferromagnetically coupled to their neighbors) at high doping requires 3$d_{3y^2-r^2}$ occupancy.  Because of the symmetry of these different orbitals the former, 3$d_{x^2-y^2}$ occupancy, will result in oblate (two short and four long bonds)  and the latter, 3$d_{3y^2-r^2}$ occupancy, in prolate (two long, four short bonds) JT-distorted octahedra.  Alignment of the long-bonds in the plane along the $b$-axis naturally explains the observed orthorhombic symmetry.\cite{ling;prb00}
In addition, a theoretical model based on the two $e_g$ orbitals by Okamoto~\etal\cite{okamo;prb01} also suggests the stabilization of 3$d_{x^2-y^2}$ and 3$d_{3y^2-r^2}$ orbitals in type A and type C/C$^*$ magnetic phases respectively.

One outstanding question is the origin of the wide spin disordered region ($0.66 \leq x \leq 0.74$).
Because of the change in symmetry of the occupied orbitals a transition from A- to C/C$^*$-type order
must be first order.  Presumably in the spin disordered region a competition exists between these
two magnetic orders that frustrates the system preventing magnetic long-range order from forming.
It would be interesting to study the local spin correlations to verify this, but lack of single crystals has, thus far, prevented such studies.
However, the local structure can be studied straightforwardly using the atomic pair distribution function (PDF) analysis of neutron powder diffraction data\cite{egami;b;utbp03} and the local magnetism can be inferred from this through application of the GK rules.

In the spin-disordered region the crystallographic structure is metrically tetragonal. This would be observed both if the {\it local} structure is tetragonal and also if it is locally orthorhombic but the locally orthorhombic domains are disordered  along the $a$ and $b$ directions.  For example, this would occur if the long bonds of JT-distorted octahedra are randomly arranged along $a$ and $b$.  The PDF
method could distinguish these two cases.  Similarly, if
\mn{3}\ and \mn{4}\ ions are spatially disordered their presence will be more apparent from
a local structural study.\cite{billi;prl96}

We applied Atomic Pair Distribution Function (PDF) analysis of neutron powder diffraction data to search for the presence of JT distorted MnO$_6$ octahedra in the doping range $0.54 \leq x \leq 0.80$ at low temperature.
Being a high resolution local structure probe, PDF technique has proved to be capable of resolving different levels of  MnO$_6$ octahedra JT distortions in cubic perovskite manganites.\cite{billi;prl96}
The advantage of this technique is that both Bragg  and diffuse scattering intensities are used, reflecting both long and short range structural correlations.  This enables us to study local structures contained in diffuse scattering found underneath and between the Bragg peaks.\cite{egami;b;utbp03}

The results indicate a gradual change of the local structure with doping rather than an abrupt phase transition as seen in the average structure. Local orthorhombicity is evident as early as $x= 0.60$ where the average structure is clearly tetragonal. This supports the idea that the sample is inhomogeneous on the nano-scale with 3$d_{x^2/y^2-r^2}$ symmetry JT distorted \mn{3}\ octahedra coexisting with undistorted \mn{3}\ and \mn{4} octahedra.  The number
of JT distorted octahedra varies smoothly with doping.

\section{EXPERIMENTS}
Finely powdered samples of \lsmo\ ($x=0.54$, 0.60, 0.64, 0.66, 0.68, 0.70, 0.72, 0.76, 0.78, 0.80) were
synthesized at Argonne National Laboratory (ANL). The synthesis method is described elsewhere.\cite{millb;cc99}
All samples were characterized using x-ray diffraction and susceptibility measurements. The oxygen
content was verified by measuring the $c-$axis parameter and was found to fall on the expected curve for
stoichiometric samples.

\begin{figure}[tb]
\includegraphics[width=0.46\textwidth, keepaspectratio=1]{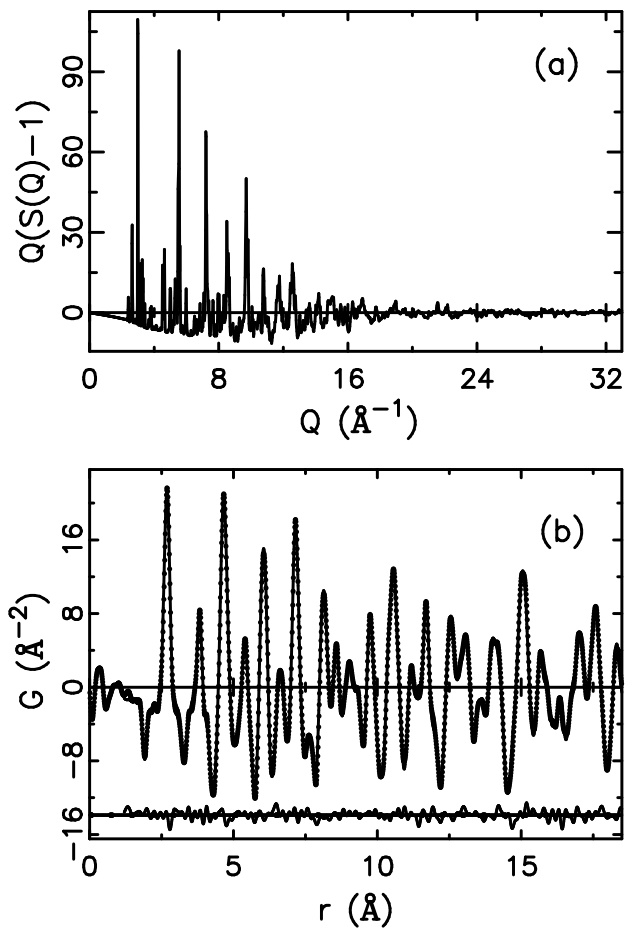}
\caption{ (a) The experimental reduced structure function $F(Q)=Q*(S(Q)-1)$ of \lsmo\ at $x=$0.80.
                (b) The experimental $G(r)$ obtained by Fourier transforming the data in (a)
                (solid dots) and the calculated PDF from refined structural model (solid line).
                      The difference curve is shown offset below. 
\label{fig;fqgr}}
\end{figure}

Neutron powder diffraction measurements were carried out on the Special Environment Powder
Diffractometer (SEPD) at the Intense Pulsed Neutron Source (IPNS) at ANL. The samples  of
about $7.0$~g were sealed in cylindrical vanadium cans with helium exchange gas.
Data were collected at $4$~K for all the compounds using a closed cycle helium refrigerator.
The $x=0.64$ and $x=0.68$ samples were measured four months after the others.
Standard corrections were made to the raw data to account for experimental effects such as
detector dead time and efficiency, background, sample absorption, multiple scattering to obtain the
normalized total scattering structure function, $S(Q)$, where $Q$ is the magnitude of the scattering vector.  These procedures are described in
detail elsewhere.\cite{egami;b;utbp03}  All corrections were carried out using the program
PDFgetN.\cite{peter;jac00}
The PDF, $G(r)$, is obtained by a Fourier transformation according to \gr .
The PDF gives the probability of finding an atom at a distance $r$ away from another atom. An
example of the PDF from \lsmo\ ($x=0.80$) at 4~K is shown in Fig.~\ref{fig;fqgr}(b)
with the diffraction data in the form of $F(Q)=Q[S(Q)-1]$ in Fig.~\ref{fig;fqgr}(a). Superimposed
on the PDF is a fit to the data of the average  structure model using the profile fitting least-squares
regression program, PDFFIT.\cite{proff;jac99}
The $S(Q)$ data were terminated at $Q_{max}=33.0$~\AA\sp{-1}. This is a reasonable value for $Q_{max}$ in
typical PDF measurements on SEPD. Uncertainties at the level of $\sigma$ are drawn as dashed lines on
the difference curves.

Peaks in $G(r)$ represent the probability of finding pairs of atoms separated by the distance-$r$,
weighted by the product of the corresponding atom pair's scattering lengths.
In a perfect crystalline La$_{2-2x}$Sr$_{1+2x}$Mn$_2$O$_7$ structure, the nearest neighbor
distance comes from the 6 equidistant Mn-O bond lengths in one MnO$_6$ octahedron, corresponding
to the first peak in PDF at about 1.94 \AA . This is negative due to Mn atom's negative neutron scattering length.
Peaks at higher-$r$ generally contain contributions from more than one unresolved pair.
The peak  at 2.72 \AA\ is dominated by high multiplicity O-O correlations, though it also contains a
contribution from La/Sr-O correlations.
The decomposition of multiple contributions is handled by a real space Rietveld refinement program:
PDFFIT,\cite{proff;jac99} with which a structural model can be obtained without the constraints
posed by space group symmetries.
Therefore, both local structural and average structure analysis can be performed on the same data set.

\section{RESULTS}

\subsection{Model independent analysis}
Low-$r$ PDF peaks directly reflect the local structural details, and disorder in the local structure  can cause excess peak broadening,\cite{billi;prl96} extra shoulders\cite{louca;prl98} and even split peaks.\cite{proff;prb99}
The prolate JT distorted octahedra of \mn{3} ions result in four short Mn-O bonds in the range $ 1.92$--1.97~\AA\  and two long bonds at 2.16~\AA .\cite{elema;jssc71,rodri;prb98,proff;prb99}  In the cubic manganites, in the absence of disorder, these are easily resolved in the PDF.\cite{proff;prb99}  With doping the loss of orientational order of the orbitals quickly suppresses the coherent JT distortion
and the average structure changes from orthorhombic to rhombohedral. However, the presence of
fully JT distorted octahedra is evident in the local structure, though the peak in the PDF from the long-bonds is not resolved and is evident only as a broad shoulder.\cite{billi;prl96,billi;prb00,billi;b;imsadoceo03}

\begin{figure}[tb]
\includegraphics[width=0.46\textwidth, keepaspectratio=1]{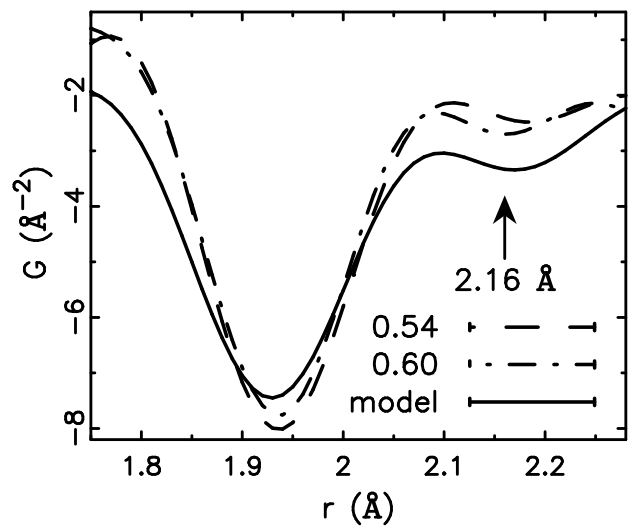}
\caption{Two dashed lines show the experimental PDFs of doping $x$
at 0.54 and 0.60. While the model PDF shown as solid line is
calculated assuming 46\% prolate octahedra (two long Mn-O bonds at
2.16 \AA, four short ones at 1.935 \AA) mixed with 54\% normal
octahedra (6 Mn-O bonds at 1.935 \AA).
\label{fig;expmod2}}
\end{figure}

\begin{figure}[tb]
\includegraphics[width=0.46\textwidth, keepaspectratio=1]{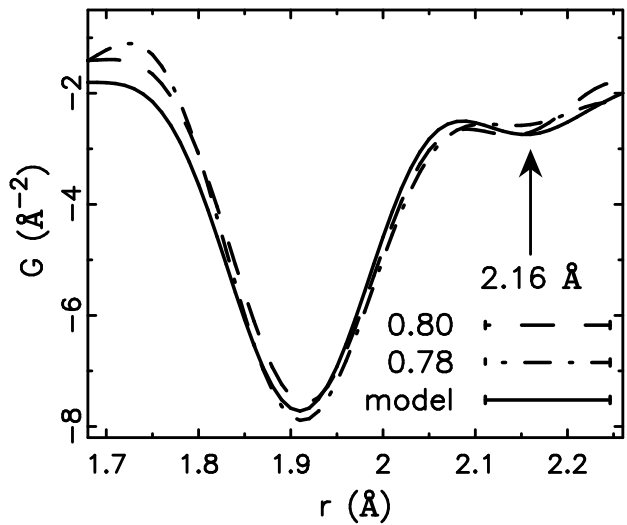}
\caption{Two dashed lines show the experimental PDFs of doping $x$ at 0.78 and 0.80. While the model PDF shown as solid line is calculated assuming 20\% prolate octahedra (2 long Mn-O bonds at 2.16 \AA, 4 short ones at 1.935 \AA) mixed with 80\% normal octahedra (6 Mn-O bonds at 1.935 \AA).
\label{fig;expmod1}
}
\end{figure}

We first investigated the PDFs from these layered, doped, manganites to search for qualitative evidence for the existence of
long $r=2.16$~\AA\ bonds.  This is shown in Figs.~\ref{fig;expmod2} and~\ref{fig;expmod1}.
Figure ~\ref{fig;expmod2} shows the experimental PDFs at $x=$0.54 and 0.60 with a
calculated PDF from a model assuming all Mn$^{3+}$O$_6$ octahedra are prolate. In
this case the number of \mn{3}\
octahedra is determined from the doping and of these, two out of six Mn-O bonds are set to 2.16~\AA .
The clear discrepancies between experiments and model rule out the existence of fully JT
distorted prolate MnO$_6$ octahedra in the type A AFI phase.  As we discuss elsewhere, this is probably due
to the fact that electrons are largely delocalized in the planes (though not perpendicular to them)
in this region of the phase diagram at low-T,\cite{qiu;unpub03} in analogy with the situation in
the CMR region of the cubic manganites.\cite{billi;prl96,billi;prb00,billi;b;imsadoceo03}

In the type C/C$^*$ orthorhombic phase, we expect e$_g$ electrons to stay in 3$d_{3y^2-r^2}$ orbitals, and therefore the 2.16 \AA\ long bonds are expect to be present. In this case the
number of \mn{3}\ sites, and therefore the number of long-bonds, is rather small.  Nonetheless,
there is rather good agreement between the prediction of the simple model and the data in the
region around $r=2.16$~\AA .
Figure ~\ref{fig;expmod1} shows the experimental PDFs at $x=0.78$ and 0.80 with the model-PDF. Because of the small number of long bonds (6.7\% at $x=0.8$) this result is not conclusive evidence supporting the existence of these long bonds, though the data are consistent with their presence.

\begin{figure}[tb]
\includegraphics[width=0.46\textwidth, keepaspectratio=1]{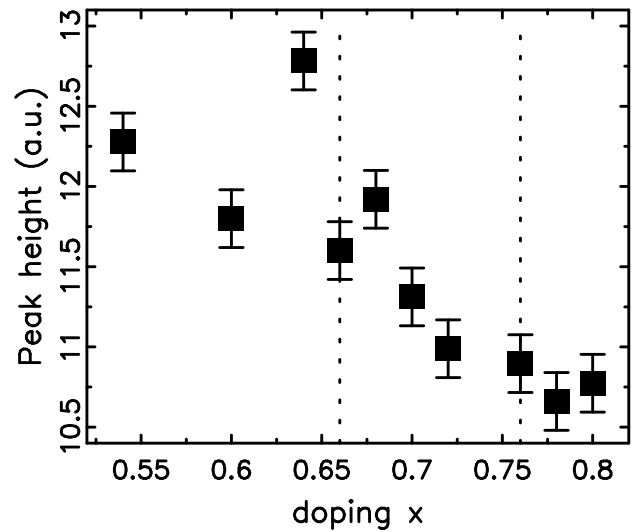}
\caption{ Solid squares show the magnitudes of the heights of the
first Mn-O peak around 1.94 \AA.  Vertical dotted lines indicate
positions of magnetic phase transitions from type-A to spin
disordered to type C/C$^*$.
\label{fig;pkheight}}
\end{figure}

The PDF peaks represent the bond length {\it distributions} in the material. The proposed MnO$_6$ octahedral
shape change of the \mn{3} octahedra, from normal to prolate JT distorted with increasing doping, induces more
local structural distortion and thus would cause the low-$r$ PDF peak to broaden.
An increase in
disorder in the local structure will result in this peak
broadening, and therefore lowering, with doping as observed.
In Fig.~\ref{fig;pkheight} we show the change of the peak height, inversely related to the peak width, of the
first PDF peak around 1.935~\AA\ with doping.    The $x=0.64$ and 
0.68 samples lie above the others.  These samples were measured at a later 
date and evidently the systematic errors have not been perfectly reproduced. Note
that the data were all collected at 4~K so no temperature
broadening effects are expected.  Also, increasing the doping in
this highly doped region moves the composition towards the pure
stoichiometric end-member and so dopant ion induced disorder
coming from the alloying is decreasing with increasing
doping.  The observation of an increase in disorder with
increasing doping, in this peak that is highly sensitive to the
Mn-O bonds, is therefore strong evidence that an electronically
driven change is occurring in the Mn-O octahedral shape.

The smooth evolution of the peak heights (Fig.~\ref{fig;pkheight})
suggests that the local structural changes occur continuously with
changes of doping concentration $x$, in contrast to what is
observed in the average structure.  The change in global symmetry
from tetragonal to orthorhombic at $x=0.76$ is presumably related
to a transition of the JT long-bonds from being randomly oriented
to having a net orientation along $b$.

\subsection{Structural Modelling}
Structural modelling gives a more quantitative picture of the
local structure than the qualitative analysis described above. Two
structural models were fit based on the tetragonal and
orthorhombic crystallographic models.\cite{ling;prb00}  It is
worth noting here that any constraint by the space group and
symmetry during average structure analysis can be relaxed in our
real space full profile modelling. Additionally, we can add any kind of
constraints based on physical reasons. It was found that the
best agreement was found for doping $x \geq  0.60$ when the
tetragonal symmetry is relaxed to orthorhombic. In the case
of $x=0.60$ the improvement in fit of the orthorhombic model is
barely significant and in this case we cannot unambiguously assign 
the local symmetry as orthorhombic. In the following, only
results from the relaxed orthorhombic symmetry are reported.

\begin{figure}[tb]
\includegraphics[width=0.46\textwidth, keepaspectratio=1]{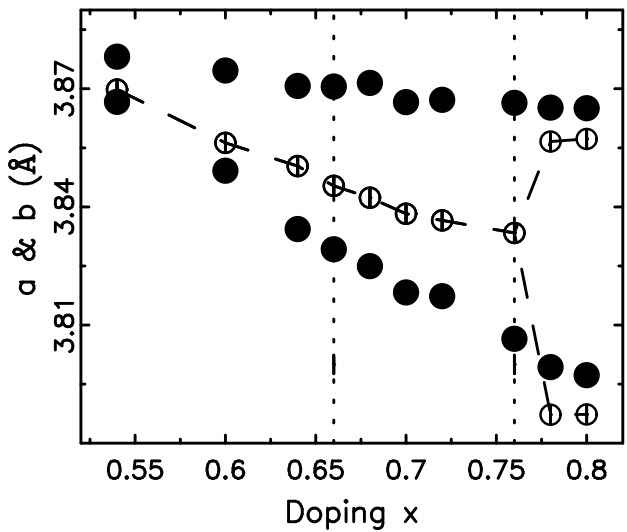}
\caption{ Solid circles are the in-plane lattice constants $a$ and
$b$ from PDF refinements with orthorhombic models, while open
circles are from Rietveld refinements on the same data with
tetragonal and orthorhombic models in $0.54 \leq x \leq 0.76$ and
$0.78 \leq x \leq 0.80$, respectively.  Vertical dotted lines
indicate positions of magnetic phase transitions from type-A to
spin disordered to type C/C$^*$.
\label{fig;latab}}
\end{figure}

The two in-plane lattice constants $a$ and $b$ are shown in
Fig.~\ref{fig;latab} together with the lattice constants obtained
from Rietveld refinement using GSAS.\cite{larso;unpub87} The two dashed
vertical lines show the phase transition from type-A AFI to spin
disordered and tetragonal to orthorhombic (which is almost coincident with 
the magnetic transition from spin disordered to type C/C$^*$ AFI), respectively.
The PDF refinements suggest the structure
is already locally orthorhombic as early as $x= 0.60$, while the
sharp crystallographic phase transition occurs around $x=0.76$.
This could be explained if JT distorted MnO\sb{6} octahedra are
beginning to appear on \mn{3} sites around $x=0.60$, but the long
bonds lie along the $a$ and $b$ axes randomly.

\begin{figure}[tb]
\includegraphics[width=0.46\textwidth, keepaspectratio=1]{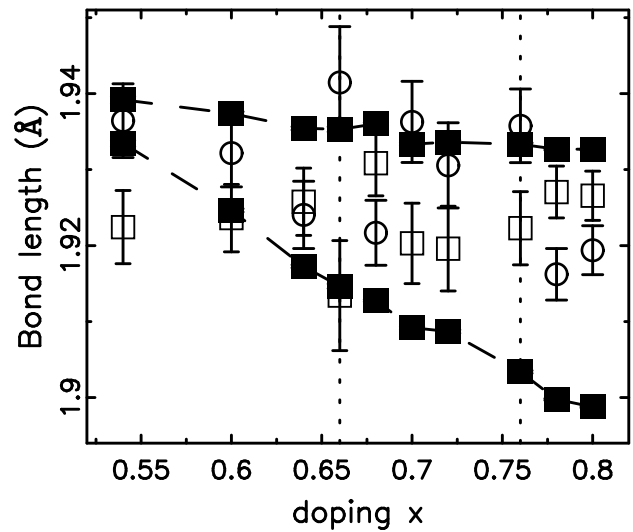}
\caption{ Solid squares are the two in-plane (ab) Mn-O bond
lengths. The bond length between Mn and the out of plane O atom is
shown as open squares, while the bond length between Mn and the
intra plane O atoms is shown as open circle.  Vertical dotted
lines indicate positions of magnetic phase transitions from type-A
to spin disordered to type C/C$^*$.
\label{fig;mnoblen}}
\end{figure}

It is important to determine whether the JT long-bonds that appear
at $x\ge 0.60$ lie in the plane, perpendicular to the plane, or are
distributed between these possibilities.  This can be studied by
looking at the refined values of the Mn-O bond lengths. The four
different Mn-O bond lengths  within one MnO$_6$ octahedron, determined by PDFFIT, are
shown in Fig.~\ref{fig;mnoblen}.
What is clear from this Figure is that over this doping range
there is no clear trend in the apical (perpendicular) bonds.
   This implies that
the observed increase in the Mn-O bond-length distribution with
doping is coming primarily from the in-plane bonds.  The JT
distorted \mn{3}\ ions that appear with doping are predominantly
locating their long-bonds in the plane. The electronic states along
$c$ remain largely unchanged with doping, and little or no
charge transfer occurs between in-plane and out of plane.

\begin{figure}[tb]
\includegraphics[width=0.46\textwidth, keepaspectratio=1]{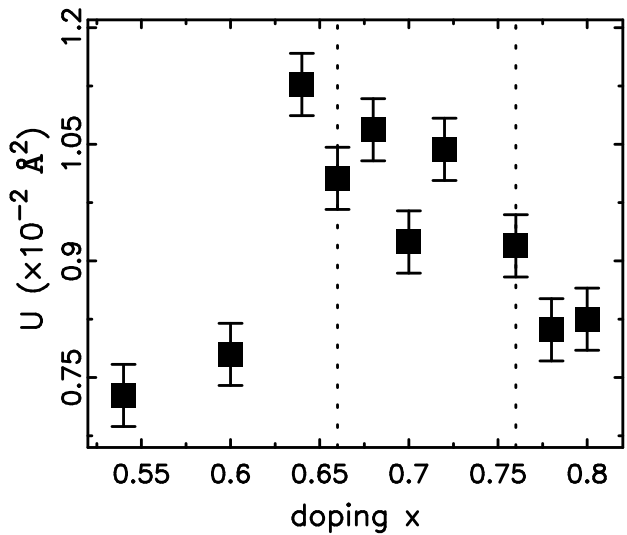}
\caption{Thermal displacement factor of the in-plane O atoms along
the direction of Mn-O bonds.  Vertical dotted lines indicate
positions of magnetic phase transitions from type-A to spin
disordered to type C/C$^*$.
\label{fig;otherm}}
\end{figure}

We have observed evidence for JT long bonds lying in the plane
from the PDF refinements.  If this picture is correct we would expect to see a
response in the refined in-plane Mn and O displacement
factors. These should be small for $x<0.60$ because there is little
disorder in the structure and the models that we are using, based
on the average structure, should work well also for the local
structure. We might expect them also to be small and largely
thermal in origin for $x>0.76$ where the JT distorted orbitals are
ordered along the $b$ axis.  In the spin disordered region we see
evidence in the local structure for significant numbers of JT
distorted  \mn{3}\ ions that are not fully ordered. By allowing
the local structure to be orthorhombic much of this disorder will
not show up in PDF derived displacement factors.  However, it is
interesting to note that there is a peak in the value of the
planar Mn-O displacement parameters in this region, as shown in
Fig.~\ref{fig;otherm}.

\section{DISCUSSION}
In a separate article we present the evidence supporting the fact
that, at low temperature at $x=0.54$, the $e_g$ electrons
associated with \mn{3}\ ions are delocalized in the plane.  At
this point there are no JT distorted MnO\sb{6} octahedra and the
local structure agrees with the average structure.  When the
C/C$^*$-type antiferromagnetism appears around $x=0.76$, coincident
with a global orthorhombic distortion, it seems clear that JT
distorted MnO\sb{6} octahedra have appeared with the
3$d_{3y^2-r^2}$ orbitals occupied.  The main result from the
current work is the observation that, in the local structure,
 the crossover between these
two behaviors happens continuously with doping and is not abrupt
as it is in the average structure.  The structure is locally
orthorhombic as early as $x=0.60$ suggesting the presence of JT
distorted MnO\sb{6} octahedra that are orientationally disordered
within the plane.  From this work we cannot tell if this disorder
is static or dynamic.

This picture could qualitatively explain the spin-disordered
region since, from the GK rules, local magnetic correlations will
randomly fluctuate between ferromagnetic and antiferromagnetic
from site to site. 

More detailed consideration of this model suggests that the sample
is likely to be nano-phase segregated in this region since we
believe the low-doping end-member (the $x=0.54$ sample) has its
$e_g$ electrons delocalized in the plane.\cite{gray;unpub03}  For
this to make sense, delocalized clusters with locally A-type
magnetic correlations must persist in this spin disordered region.
Since there is no evidence of macroscopic phase-separation these
clusters are likely to be nano-scale.  Presumably they coexist
with nanoscale regions of the sample where the $e_g$ electrons are
localized as \mn{3}\ ions with a local JT distortion.  With
increasing doping the number of these localized \mn{3}\ sites
first increases as the proportion of the sample in the localized
state increases at the expense of the delocalized state.  Once the
entire sample has transformed to the localized state, with
increasing doping the number of \mn{3}\ sites will decrease as
($1-x$) in the normal way.  This is apparent from the decreasing
orthorhombicity that is evident for $x>0.80$.\cite{ling;prb00}

\section{conclusions}
Based on PDF results, we suggest that the local structure of 
\lsmo\ evolves smoothly as a function of doping at low temperature in the
region of the phase diagram ${0.54 \leq x \leq 0.80}$.  The material evolves
smoothly from being locally tetragonal at $x=0.54$ to having a well established 
orthorhombicity at $x=0.80$.  The local and global structures agree well at these
end-points.  However, in between, and associated with the spin disordered region
of the phase diagram, the local structure appears orthorhombic even though the 
material is metrically tetragonal.  These results can be reconciled if JT distorted
MnO\sb{6} octahedra exist with their long-bonds lying in the plane but disordered
along the $a$ and $b$ axes.  We have discussed that these results are consistent with
the presence of inhomogeneities resulting in a coexistence of delocalized and localized 
electronic states, possibly due to nano-scale phase separation, 
in this intermediate region of the phase
diagram, into regions that have the characteristics of the two end-members at $x=0.54$ and 
$x=0.80$ respectively. We have argued that such a nano phase separation into disordered and possibly fluctuating A-type and C/C$^*$-type magnetic domains may explain the frustrated magnetism in this region.  Making certain assumptions we have quantified the evolution 
of the phase separation with doping.

\acknowledgments
We are thankful to Simine Short for providing valuable help with
the neutron diffraction data collection. The work at MSU was supported
 by NSF through grants DMR-0075149, and at ANL by the US Department of Energy,
Office of Science, under Contract No. W-31-109-ENG-38. The IPNS at Argonne
National Laboratory is funded by the US Department of Energy under
contract W-31-109-ENG-38. 



\bibliographystyle{\bibpathsjb{aip_simon}}


\end{document}